\documentclass[prb,aps,twocolumn,showpacs,floatfix]{revtex4}

\usepackage{amsmath,amssymb}
\usepackage{graphicx}
\usepackage{version}
\DeclareGraphicsExtensions{.eps,.ps}

\newcommand{\br}{{\bf r}}
\newcommand{\bn}{{\bf n}}

\newcommand{\beqa}{\begin{eqnarray}}
\newcommand{\eeqa}{\end{eqnarray}}

\begin{document}


\title
{Bilayer graphene Origami: curvature-induced p-n junctions}
\author{Yogesh N. Joglekar}
\affiliation{Department of Physics, Indiana University-Purdue
University Indianapolis (IUPUI), Indianapolis, Indiana 46202, USA}
\author{Avadh Saxena}
\affiliation{Theoretical Division, Los Alamos National Laboratory, Los Alamos, 
New Mexico 87545, USA}
\date{\today}

\begin{abstract}
A massive quantum particle on a two-dimensional curved surface experiences a 
surface-geometry induced attractive potential that is characterized by the 
radii of curvature at a given point. With bilayer graphene sheets and 
carbon nano-ribbons in mind, we obtain the geometric potential $V_G$ for 
several surface shapes. Under appropriate conditions that we discuss in 
detail, this potential suppresses the local Fermi energy. Therefore, we 
argue that in zero band-gap materials with a quadratic band structure, it will 
create p- and n-type regions. We discuss the consequences of this result, 
and suggest that surface curvature can provide a novel avenue to create p-n 
junctions and, in general, to control local electronic properties in carbon 
nano-ribbons and bilayer graphene sheets. 
\end{abstract}

\maketitle


\noindent{\it Introduction:} 
The classical dynamics of a particle subject to (holonomic or non-holonomic) 
constraints is well-understood. In Lagrangian formulation, these constraints 
can be represented by an effective potential.~\cite{goldstein} In quantum 
theory, the potential representing constraints plays an instrumental role in 
determining the quantized spectrum of the system.~\cite{griffith} Indeed, 
low-dimensional systems are realized by constraints that classically restrict 
the motion of the particle in one or more directions. For example, in 
two-dimensional electron (or hole) systems (2DES), typically realized in 
III-V semiconductor heterojunctions~\cite{sundaram} the barrier potential at 
the heterojunction interface, created by differing electron affinities, leads 
to a clear separation of energy scales for the motion in the 2D plane and 
motion along the direction normal to the plane. It allows one to separate 
the particle wavefunction into normal and surface components and, at low 
energies, focus solely on the surface component.~\cite{ando}

The quantum properties of a non-relativistic particle 
constrained~\cite{jensen} on an arbitrary orientable surface were first 
investigated by da Costa.~\cite{costa,caveat1} He showed that apart from the 
necessity of expressing the Laplace-Beltrami operator $\nabla^2$ in terms of 
the curvilinear surface co-ordinates, the geometry of the surface also leads 
to an attractive geometric potential 
$V_G(q_1,q_2)=-(\hbar^2/8m)(\kappa_1-\kappa_2)^2$ where $m$ is the mass of 
the particle, $(q_1,q_2)$ denote surface co-ordinates, 
and $\kappa_1,\kappa_2$ are the two position-dependent principal curvatures 
of the surface. Note that this potential is purely a result of particle 
confinement, and therefore is independent of the electric charge of the 
particle; it is therefore the same for electrons and holes. This result is 
applicable in the limit $W\kappa\rightarrow 0$ where $W$ is the thickness 
of the ``two-dimensional'' surface and $\kappa$ is the surface curvature. 
$W$ will correspond to the width of the quantum well in 2DES. The geometric 
potential is small near a surface maximum or minimum, and is large near 
saddle regions where $\kappa_1\kappa_2<0$. In particular, we note that for 
surfaces with a constant curvature (spherical sections or a plane) the 
geometric potential is equal to zero, $V_G=0$. The bound states that 
appear due to the geometric potential~\cite{goldstone} and the resultant 
resonance microwave absorption~\cite{carini} have been investigated in 
twisted~\cite{clark} or bent (quasi one-dimensional electron) 
waveguides.~\cite{timp} The effect of curvature on spin-orbit coupling has 
been studied in the case of nanospheres and nanotubes.~\cite{magarill} 
More recently, the geometric-potential induced charge separation in 
helicoidal ribbons has been analytically explored.~\cite{atanasov} However, 
absent a truly two dimensional system that can be (easily) bent, the effect of 
geometric potential on the electronic band structure has been justifiably 
ignored.

In this paper, we point out that the geometric potential provides a hitherto 
unexplored avenue to locally change the carrier polarity in zero-gap 
materials. We predict that it will be possible to create p-n or n-p-n 
junctions purely from a suitable surface geometry. Bilayer graphene 
provides an ideal realization of a zero-gap material with chiral electrons 
and holes that have identical mass. Graphene nano-ribbons (GNR), recently 
created by unzipping single and multi-walled carbon nanotubes~\cite{gnr} may 
provide another promising candidate that can be used to explore the 
curvature effects in a gapless semiconductor. Therefore, in this paper we 
primarily focus on bilayer graphene. In the next section, we obtain the 
geometric potential for two generic cases: first a surface $z=f(x)$ that is 
isometric to a plane and second a catenoid that is {\it not} isometric to a 
plane. In the following section, we show that at present carrier densities 
$n_{2D}$, carriers in bilayer graphene, with its Bernal-stacked hexagonal 
lattice structure, can be treated as massive particles on a curved surface. 
We conclude the paper with a qualitative discussion about experimental 
consequences and speculation regarding the geometric potential for 
monolayer graphene with massless chiral carriers. 

Monolayer and bilayer graphene have been studied in great detail in recent 
years. P-N junctions in monolayer graphene are fabricated by local 
top-gating~\cite{pnmarcus} and with an air-bridge top 
gate.~\cite{pnairbridge} Bilayer graphene has been explored for its electric  
field tunability.~\cite{taisuke,oostinga} The effect of random surface 
curvature 
(ripples) on electronic properties in monolayer graphene is primarily 
described via gauge potentials~\cite{gaugeguinea}; however, the effect of 
geometric potential in monolayer and bilayer graphene has not been 
investigated. In {\it monolayer} graphene near the neutrality point, disorder 
induced electron-hole puddles and the resulting array of p-n junctions is 
observed.~\cite{martinyacobi} However, in this case, the origin of the 
disorder and the role, if any, played by monolayer surface curvature is 
unclear.~\cite{martinyacobi} In particular, the topic of p-n junctions made 
with bilayer graphene (that we propose in this paper) has been scarcely 
investigated. 

In this paper, we argue that due to its essentially two-dimensional 
nature ($W\kappa\rightarrow 0$), gapless quadratic band structure, and 
exceptional material strength, bilayer graphene presents an excellent 
candidate for exploration of two-dimensional systems with curvature.

\begin{figure}[thbf]
\includegraphics[width=0.6\columnwidth]{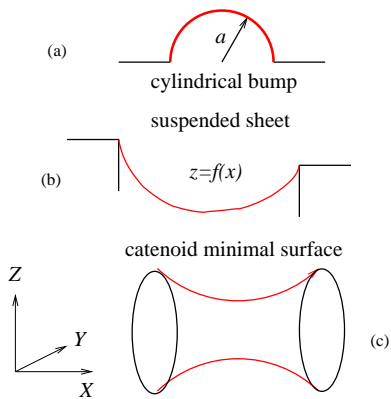}
\caption{(Color online) Schematic shapes of various curved surfaces 
considered in this paper. The top two, a sheet with a cylindrical bump with 
radius $a$ (a) and a sheet suspended between two mesas (b), have only 
one nonzero principal curvature $\kappa(x)$ and are isometric to a plane. 
The surface at the bottom, a catenoid (c), has two nonzero principal 
curvatures at each point $\kappa_1(x,y)\kappa_2(x,y)<0$.} 
\label{fig:schematic}
\end{figure}

\noindent{\it Continuum Model:} 
Let us consider a non-relativistic particle of (effective) mass $m$ 
constrained to a surface given by $z=f(x)$ (See Fig.~\ref{fig:schematic}). 
We characterize a point on the two-dimensional surface embedded in the 
three-dimensional space by $\br=(x,y,f(x))$. The geometric potential is 
obtained from the two fundamental forms associated with the 
surface.~\cite{costa} The first fundamental form or, equivalently, the 
(diagonal) metric tensor~\cite{boothby} for the surface is given by 
$g_{ij}=\partial_i\br\cdot\partial_j\br=
{\mathrm{dia}}(1+f'^2(x),1)$ where $i,j=x,y$ and $f'=df/dx$. The second 
fundamental form $h_{ij}=-\partial_i\br\cdot\partial_j\bn$ characterizes the 
change in the surface-normal $\bn$ along the surface~\cite{boothby} and has 
only one nonzero element, $h_{xx}(x)=-f''(x)/[1+f'(x)^2]^{3/2}$ where 
$f''=d^2f/dx^2$. Principal curvatures of the surface are eigenvalues of the 
matrix~\cite{costa} $\alpha=-h g^{-1}$ and, in this case, they are given by 
$\kappa_1(x)=f''(x)/\sqrt{1+f'(x)^2}$ 
and $\kappa_2=0$. The zero eigenvalue reflects the translational invariance 
of the system in the $y$-direction, and the zero Gaussian curvature 
$K=\kappa_1\kappa_2=0$ everywhere on the surface implies that it is isometric 
to a plane.~\cite{boothby} Thus, for such a surface, the geometric potential 
is given by $V_G(x)=-(\hbar^2/8m)[f''(x)/\sqrt{1+f'(x)^2}]^2$. 

We remind the reader that the potential is purely quantum, can be different 
for isometric surfaces (for example, a plane and a cylinder), and is always 
attractive. The Hamiltonian for the constrained particle is 
$\hat{H}=\hat{H}_0 +\hat{V}_{G}$ and the kinetic contribution is given by 
\begin{eqnarray}
\label{eq:ke1}
\hat{H}_0 & = &-\frac{\hbar^2}{2m}\frac{1}{\sqrt{g}}\partial_i\left[\sqrt{g}
(g^{-1})_{ij}\partial_j\right]\\
\label{eq:ke2}
& = &-\frac{\hbar^2}{2m}\left[\frac{d^2}{ds^2}+\frac{d^2}{dy^2}\right] , 
\end{eqnarray}
where $g=\det g_{ij}$, and $s(x)=\int_0^{x}du\sqrt{1+f'(u)^2}$ is the 
arc-length along the surface. In terms of the surface co-ordinates $(s,y)$, 
the geometric potential term becomes
\begin{equation}
\label{eq:vcosta}
V_G(s(x),y)=-\frac{\hbar^2}{8m}\left[\frac{s''(x)}{s'(x)^2\sqrt{s'(x)^2-1}}
\right]^2.
\end{equation}
Note that for a cylindrical bump of radius $a$ (Fig.~\ref{fig:schematic}a), 
we recover~\cite{costa} a constant geometric potential $V_G=-\hbar^2/8ma^2$. 
This attractive geometric potential is independent of the sign of the 
curvature: a bump, a trough, or combinations thereof lead to the same 
constant attractive potential. For a suspended sheet 
(Fig.~\ref{fig:schematic}b),~\cite{caveat2} $f(x)=a\left[1-\cosh(x/a)\right]$ 
where $a$ sets the length-scale for the catenary, the geometric potential is 
given by 
\begin{equation}
\label{eq:catenary}
V_G(s(x),y)=-\frac{\hbar^2}{8ma^2}\left[\frac{a^2}{a^2+s^2}\right]^2.
\end{equation}
Lastly, we consider the particle confined to a catenoid (Fig.~\ref{fig:schematic}c) 
parameterized by two dimensionless surface co-ordinates $(u,v)$ where 
$x=a\cosh(u)\cos(v)$, $y=a\cosh(u)\sin(v)$ and $z=au$. Note that perfect (bilayer) 
graphene sheets are planar and, as such, they cannot be mapped onto a surface 
with nonzero Gaussian curvature. However, fullerenes or dislocations in the 
honeycomb lattice can lead to surfaces with nonzero Gaussian curvature. A 
straightforward calculation gives the geometric potential 
\begin{equation}
\label{eq:catenoid}
V_G(u,v)=-\frac{\hbar^2}{2ma^2}\frac{1}{\cosh^4(u)}.
\end{equation}
We note that $V_G(u,v)$ is solely a function of $u$, just as $V_G(s,y)$ in 
Eq.(\ref{eq:vcosta}) is solely a function of $s$. However, unlike in the 
former case, the kinetic energy term on the catenoid $\hat{H}_0=-
(\hbar^2\kappa_1(u)/2ma)(\partial^2/\partial u^2 +\partial^2/\partial v^2)$ 
is not separable and reflects the nonzero Gaussian curvature of the surface. 


\noindent{\it Applicability to Bilayer Graphene:}
To map a curved bilayer graphene sheet onto a quantum particle on 
a curved surface, two criteria must be satisfied: $k_F a_G\ll 1$ and 
$\kappa a_G\ll 1$ where $a_G=1.4$ \AA\, is the carbon-carbon distance in 
graphene, $k_F$ is the Fermi wavevector, and $\kappa$ is the surface 
curvature. The first criterion ensures that bilayer graphene can be treated 
in the continuum limit. In this limit, for carrier densities $n_{2D}\lesssim
3.5\times 10^{13}$ cm$^{-2}$, only the gapless valence and conduction bands 
are occupied~\cite{falko,biased} and the carriers can be approximated as 
particles with an effective mass $m^*\sim 0.03 m_e$. The 
second criterion ensures the detailed lattice structure can be ignored on the 
curvature length-scale, and because the ``thickness'' of bilayer graphene 
sheets is $W\sim 3 a_G$, it also ensures that $W\kappa\ll 1$, or that the 
bilayer graphene sheet with nonzero thickness can be considered a surface. 
The first criterion is easily satisfied at present carrier densities in 
bilayer graphene with $n_{2D}\sim 10^{12}$ cm$^{-2}$ and corresponding 
$k_F\sim 10^{-2}$\AA$^{-1}$. The second criterion, $\kappa a_G\ll 1$ is 
satisfied by carbon nanotubes with typical radii $1/\kappa\sim$ 10-30 \AA\,  
and by typical surface ripples on monolayer graphene.~\cite{ripples} When 
$\kappa\sim k_F$ 
the geometric potential $V_G$ becomes comparable to the Fermi energy 
$E_F=\hbar^2 k_F^2/2m^{*}$, whereas in ``planar'' bilayer graphene 
$\kappa\approx 0\ll k_F\ll 1/a_G$. 

Within the continuum approximation, we can consider two cases. When 
$1/a_G\gg \kappa\gg k_F$ the problem is reduced to that of a single-particle 
in an attractive external potential.~\cite{atanasov} This case is not very 
realistic since at (vanishingly) low carrier densities, disorder effects 
dominate the physics.~\cite{martinyacobi,shaffiq} On the other hand, when 
$\kappa\sim k_F\ll 1/a_G$, the geometric potential will reduce the local 
Fermi energy. In particular, if we consider graphene with n-type 
carriers in the flat region ($E_{F0}>0$), the geometric potential will lower 
the Fermi energy from the conduction band to the valence band, 
$E_F(q_1,q_2)=E_{F0}-|V_G(q_1,q_2)|<0$. Note that as long as the 
geometric potential is constant or slowly varying (as would be the case for 
a cylindrical bump), this result is robust and is independent of any other 
quantum numbers, such as chirality or electric charge, that the carriers 
may have. Therefore, {\it we predict that in such a configuration, the 
curved regions will have p-type carriers with a natural p-n junction created 
between the curved and flat regions}. For the surface $z=f(x)$, the geometric 
potential will create p- and n-type strips where the location, size, and 
polarity of each strip is determined by the local Fermi energy $E_F(x)$. 

Bilayer graphene is distinguished from its semiconductor counterparts by the 
chiral nature of its carriers that leads to zero transmission probability 
for normal incidence at a potential barrier in spite the zero gap between 
the electron and hole bands.~\cite{klein} Thus, the chirality of the 
carriers in bilayer graphene will only affect the transparency (i.e. the 
transmission amplitude as a function of incidence angle) of the p-n 
junctions created by the geometric potential.~\cite{caveat3} Such p-n 
junctions can be created by either depositing bilayer graphene on a patterned 
substrate with requisite bumps and troughs with radii $R\sim$ 100 \AA\, or by 
suspending it 
across a narrow channel. A local probe~\cite{mechgraphene} may also offer an 
alternative way to induce controlled curvature in a suspended bilayer 
graphene sheet. We emphasize that our prediction is valid for any 
two-dimensional zero-gap semiconductor with appropriate curvature, although 
bilayer graphene and GNRs are the most promising candidates.


\noindent{\it Discussion:} In this paper we have proposed that bilayer 
graphene on a patterned substrate provides a novel avenue to create a p-n 
junction. Our prediction provides a direct way to experimentally probe the 
effect of geometric potential that has been, hitherto, neglected in 
two-dimensional systems. 

The origin of the geometric potential $V_G$ is well-understood for a 
non-relativistic massive particle.~\cite{costa} In case of a planar 2DES, a 
lattice model naturally leads to carriers with an effective mass. However, a 
lattice-model generalization of the geometric potential is, to our knowledge, 
an open question. The primary difficulty in addressing such a question is 
that typical lattice models start with zero thickness, $W=0$, whereas the 
geometric potential arises from taking the limit $W\rightarrow 0$.

We conclude the paper with speculation about curved monolayer graphene. Since 
the geometric potential $V_G$ is obtained from 
non-relativistic Schr\"{o}dinger equation, {\it prima facie} is not 
applicable to monolayer graphene where the carrier dynamics is well-described 
by the two-dimensional Dirac Hamiltonian. Monolayer graphene with random 
curvature (ripples) is modeled using random gauge fields~\cite{gaugeguinea} 
and mean-curvature-dependent potential.~\cite{neto} We are unaware of a 
rigorous derivation of geometric potential for massless particles or for a 
lattice-model that leads to linearly dispersing bands. Dimensional analysis 
implies that for a surface $z=f(x)$, the geometric potential, if nonzero, 
must scale as $V_G\propto\hbar v_G|\kappa|$ where $v_G$ is the velocity of 
massless carriers in graphene.~\cite{min} It follows that under the 
conditions discussed in the last section, this geometric potential will 
locally change the Fermi energy $E_F(x)$ and will lead to the formation of a 
p-n junction. We remind the reader that the transparency of such a junction 
in monolayer graphene is drastically different from that in bilayer graphene 
although both are zero-gap materials.~\cite{klein} 

The test of our prediction via experiments and the generalization of the 
geometric potential to a lattice model will provide insights into the 
electronic structure of curved two-dimensional materials with zero band-gap. 
Curvature induced p-n and p-n-p junctions in bilayer graphene and the GNR 
may open up new vistas for nano-electronics.  


\noindent{\it Acknowledgments:} Y.J. thanks Sasha Balatsky for the 
opportunity to visit Los Alamos National Laboratory and acknowledges the 
hospitality of KITP, Santa Barbara (Grant No. NSF PHY05-51164) during the 
completion of this work.  This work was supported in part by the U.S. 
Department of Energy. 



\begin{thebibliography}{99}
\bibitem{caveat1} The role of curvature in the spectrum of a particle confined 
between equidistant surfaces was explored in ref.~\onlinecite{jensen}.
\bibitem{caveat2} Strictly speaking, this expression is true only for an 
infinite hanging sheet.
\end{thebibliography}
\end{document}